\documentclass[twocolumn,superscriptaddress,showpacs,aps,prb]{revtex4}
\usepackage{mathtext}
\usepackage{graphicx}
\usepackage[dvips]{epsfig}
\usepackage{amsmath}
\usepackage{slashbox}
\newcounter{fig}

\begin{document}
\title{Magneto-optics  of monolayer and bilayer graphene  }
\author{L.A. Falkovsky}
\affiliation{L.D. Landau Institute for Theoretical Physics RAS, 119334 Moscow
\\L.F. Verechagin Institute of the High Pressure
Physics  RAS,  142190 Troitsk}
\date{\today}

 \begin{abstract}
The optical conductivity of graphene and bilayer graphene in quantizing magnetic fields is studied.
Both  dynamical  conductivities, longitudinal and Hall's, are 
analytically evaluated. The conductivity peaks are explained in terms of electron transitions. Correspondences between the transition frequencies and the magneto-optical features are established using the theoretical results.  The main optical transitions obey the selection rule with $\Delta n=1$ for the Landau number $n$. 
The Faraday rotation and light transmission in the quantizing magnetic fields are calculated. The effects of temperatures and magnetic fields on the chemical potential are considered.
\end{abstract}
\maketitle

\section{Introduction}
 The most accurate  investigation of the band structure of metals and semiconductors is  a study of   the Landau levels  through experiments such as magneto-optics \cite{ST,TDD,DDS,MMD,LTP,OFM,OFS,OP,CLW} and magneto-transport \cite{KTS,LK,JZS,SOP,RM}. In magnetic fields,  the classical and quantum Hall effects are  observed, as well as the polarization rotation for transmitted    
    (Faraday's rotation) or reflected (Kerr's rotation) lights. However, the interpretation of the experimental results involves a significant degree of uncertainty, because it is not clear how the resonances
can be identified and which electron transitions they correspond to.

Comprehensive literature on the graphene family can be described in terms of the Dirac gapless fermions. According to this picture,  in graphene, there are two
bands at the $K$  hexagon vertexes of the Brillouin zone  without any gap between them, and the electron dispersion can be considered as linear  in the wide wave-vector region. For the dispersion linearity, this region should be small
compared with the size of the Brillouin zone, i.e. less than 10$^{8}$ cm$^{-1}$, providing  the small carrier concentration $n_0\ll 10^{16}$ cm$^{-2}$. Pristine graphene  at  zero temperature has no carriers, and the Fermi level should divide  the conduction and valence bands. However, undoped graphene cannot be really obtained, and so far purest graphene contains about $n_0\sim10^9$ cm$^{-2}$ of carriers. Then the following problem appears --- how do  Coulomb electron-electron interactions renormalize the linear dispersion and does graphene become an insulator  with a gap?

 Semiconductors with the gap are needed for electronic applications.  
 Investigations of the  graphene bilayer and multilayer are very popular as the  gap appears when  the bias is applied. 
   Here Slonczewski,  Weiss, and McClure (SWMC) should be mentioned because they 
 have stated the description  of a layered matter \cite{SW} with  interactions strong in a layer and  weak  between layers. The theory 
 contains several parameters which are the hopping integrals for nearest neighbors. Such a picture has been examined in many experiments \cite{BCP}.

  The theoretical  solution for the band problem in magnetic fields  often cannot be  exactly found.  A typical example is presented by  graphene layers. For bilayer graphene and graphite,
 the effective Hamiltonian is a $4\times4$ matrix giving four energy bands.
 The trigonal warping described by the small parameter $\gamma_3$ in the effective Hamiltonian  provides an evident effect. Another important parameter is the gate-tunable bandgap $U$ in bilayer graphene. 
  In this situation, the quantization problem cannot be  solved within a rigorous method. To overcome this difficulty, several methods have been proposed for approximate \cite{OP,Fa,LA,CBN,ZLB,Ko}, numerical \cite{UUU,Nak,PP,GAW}, and semiclassical quantization \cite{Falko,Dr,ZFA,OF}. 
  
  In this paper, our attention is focused on the dynamic conductivity of monolayer and bilayer graphene in the presence of a constant magnetic field in $z-$direction. We consider the collisionless limit when the electron collision rate is much less then the frequency of the electric field. Then, the accurate theoretical results can be obtained for Faraday's rotation and transmittance through graphene layers. The present paper is organized as follows. In Sec. II we recall the electron dispersion in the monolayer and bilayer graphene.  In Sec. III   we describe in detail the quantization in magnetic fields. In Sec. IV the longitudinal and Hall conductivities as well as the Faraday rotation are described. Effects of temperatures and magnetic fields  on the chemical potential are considered  in  Sec. V. Section VI contains a summary of the discussed results.
 
  \section{Electron dispersion in monolayer and bilayer graphene  }
\emph{ Electron dispersion in  graphene.}
   
The    symmetry  of the $K$ point is $C_{3v}$ with the threefold axis and   reflection planes.  This group has twofold  representation with the 
basis functions transforming  each in other under reflections and obtaining the factors $\exp{(\pm2\pi i/3)}$ in rotations. The linear momentum displacements from the $K$ point, taken as $p_{\pm}=\mp ip_x-p_y$, transform in a similar way. The effective Hamiltonian is invariant under the group transformations, and we have the unique possibility  to construct the invariant  Hamiltonian linear in the momentum as
     \begin{equation}
H(\mathbf{p})=\left(
\begin{array}{cc}
0 \,    & vp_{+} \\
vp_{-} \,& 0      
\end{array}%
\right) \,,  \label{ham0}
\end{equation}%
where $v$ is a constant of the velocity units.  The same Hamiltonian can be written using the tight-binding model. 

The eigenvalues of this matrix give  two bands 
\[\varepsilon_{1,2}=\mp v\sqrt{p_x^2+p_y^2}=\mp vp\,,\]
where the subscript $s=1,2$ numerates  these two bands  (holes and electrons). 
The gapless linear spectrum arises as a consequence of the symmetry, and  the Fermi energy coincides with the band crossing (the Dirac point) due to the carbon valence. The cyclotron mass has the form
\[m(\varepsilon)=\frac{1}{2\pi}\frac{dS(\varepsilon)}{d\varepsilon}=\frac{\varepsilon}{v^2}\,,\]
and the carrier concentration at zero temperature  
$ n(\varepsilon_F)=\varepsilon_F^2/\pi\hbar^2v^2$
is simply expressed in terms of the Fermi energy $\varepsilon_F$. 

Tuning the gate voltage, the linearity of the spectrum has been examined  in the Schubnikov--de Haas  studies \cite{EGM} with the help of the connection between the effective mass and the carrier concentration  at the Fermi level
$m(\varepsilon_F)v=\mp\hbar\sqrt{\pi n(\varepsilon_F)}$.
 The "constant"\, parameter $v$ was found to be no longer constant. At low carrier concentrations   
$n\sim10^9$ cm$^{-2}$,  it exceeds   its constant value
$v=1.05\pm0.1\times10^8 $ cm/s for  concentrations $n>
10^{11}$ cm$^{-2}$ by the factor of 3.

This is a result of  electron-electron interactions which become stronger at low carrier concentrations. The logarithmic renormalization of the velocity was found by Abrikosov and Beneslavsky \cite{AB} for the three-dimensional case
and in Refs. \cite{Mi,GGV} for two-dimensional graphene. Notice, that no phase transition  was revealed even at lowest carrier concentration. We have also to conclude that  the Coulomb interactions do not create any gap  in the graphen spectrum.
 
We recall the peculiarity of graphene conductivity in the absence of the magnetic field \cite{FV,GSC}. For the optical frequency range, when the spacial dispersion of conductivity is not significant, the intraband electron transitions make a contribution    
\begin{equation}
     \sigma ^{intra}(\omega) =\frac{2ie^2T}
     {\pi\hbar(\omega+i\tau^{-1})}
\ln{(2\cosh\frac{\mu}{2T})}
 \label{sigm}    \, ,
 \end{equation}
which has the Drude--Boltzmann form at the large chemical potential $\mu\gg T.$ 

At the zero temperature, the interband electron contribution can be presented in the simple form 
\begin{equation}\nonumber
    \sigma^{inter}(\omega) =
    \frac{e^2}{4\hbar}\left[\theta(\omega-2\mu)-\frac{i}{2\pi}\ln
    \frac{(\omega+2\mu)^2}{(\omega-2\mu)^2}
     \right]\,,
 \label{ibd} \end{equation}
 where the $\theta-$function expresses the  threshold behavior of interband electron transitions at $\omega=2\mu$. The temperature  smooths out all the singularities in this formula.
 In high frequency region
$\omega\gg(T,\mu)$, the interband transitions make the leading contribution into  conductivity 
$$\sigma(\omega) =
    \frac{e^2}{4\hbar},$$
     having the universal character independent of any material parameters. This frequency region is limited above by the band width of around 3 eV. 
     
     Making use the universal conductivity, one can calculate the light transmission through graphene \cite{FP, Kuz} in the  approximation  linear in conductivity 
\begin{equation}\nonumber
  T=1-\frac{4\pi}{c} Re\,\sigma(\omega)\cos{\theta}=1-\pi\frac{e^2}{\hbar
  c}\cos{\theta}\,,
  \label{transmis}\end{equation}
where $\theta$ is the incidence angle of light.  In excellent agreement with the theory, for the wide optical range,   several experimental groups \cite{NBG, Li, Ma} observe   the light transmission through graphene as well as bilayer graphene where the difference from unity is twice as larger.  It is exceptionally intriguing that the light transmission involves the fine structure constant  $\alpha= e^2/\hbar c $ of quantum electrodynamics having really no relations to the graphene physics. 

For the frequenciy range, where the intraband term plays the main role, the plasmon excitations are possible \cite{FV,HDS} with the dispersion 
\[
\omega =\sqrt{\kappa k},\quad \kappa =\frac{2e^{2}T}{\hbar
^{2}}\ln ({2\cosh{\frac{\mu}{2T}}}).
\]%
and relatively small damping, determined by the electron relaxation $\tau^{-1}$. The plasmon has the same dispersion, $k^{1/2}$, as the normal 2d plasmon. However, it shows the temperature dependence at low carrier concentrations, $\mu<2T$.

\emph{ Electron dispersion in  bilayer graphene. }

  Bilayer graphene has attracted much interest partly due to the
opening of a tunable gap in its electronic spectrum with   an
external electrostatic field. Such a phenomenon was predicted in
Refs.  \cite{McF,LCH} and was observed in optical studies
controlled  by applying a gate voltage
\cite{OBS,ZBF,KHM,LHJ,ECNM,NC,MLS,KCM}. 

 The Hamiltonian of the SWMC theory can be written \cite{PP,GAW} near the $ K$ points  in the Brillouin zone in the form
\begin{equation}
H(\mathbf{p})=\left(
\begin{array}{cccc}
U    & vp_{+} & \gamma_1 &\gamma_4vp_{-}/\gamma_0\\
vp_{-} & U  & \gamma_4vp_{-}/\gamma_0& \gamma_3vp_{+}/\gamma_0\\
\gamma_1  &\gamma_4p_{+}/\gamma_0 & -U  &vp_{-}\\
\gamma_4vp_{+}/\gamma_0 & \gamma_3vp_{-}/\gamma_0 &vp_{+} &-U
\end{array}%
\right)\,,   \label{hamg}
\end{equation}%
where $p_{\pm}=\mp ip_x-p_y$. 
 The nearest-neighbor hopping integral $\gamma_0\approx 3$ eV corresponds with the velocity parameter $v=1.5a_0\gamma_0 = 10^6$ m/s and the in-layer inter-atomic distance $a_0=1.415$ \AA\,. The parameters $\gamma_{3,4}\sim 0.1$ eV describe the interlayer interaction at the distance $d_0$=3.35\AA\, between   layers. 

  Hamiltonian  (\ref{hamg}) give four levels labeled by the number $s=1,2,3,4$ from the bottom. For $U=0$, the twofold degeneration  $\varepsilon_2=\varepsilon_3$ exists at  $p_x=p_y=0$, as a consequence of axial symmetry.
The parameter $U$ is included in the  bilayer Hamiltonian to describe  the gate voltage. 
At $U\neq 0$, the gap appears between the $\varepsilon_2$ and  $\varepsilon_3$, and these bands acquire the form of "mexican hat"\,.

 Two vertexes, $K$ and $K'$, in the Brillouin zone are transforming each in other under reflection. Such the reflection changes the $U$ sign. Therefore, the levels at  these vertexes  do not coincide in the presence of a gate. The levels at  $K'$ point can be obtained from levels at the $K$ point  by changing their signs.

\section{Graphene in magnetic fields}
In the presence of the magnetic field $B$, the momentum projections $p_+$ and $p_-$ become the operators with the commutation rule $\{\hat{p}_{+},\hat{p}_{-}\}=-2e\hbar B /c$. 
 We  use the relations
\[v\hat{p}_+=\omega_B\,a, \quad  v\hat{p}_-=\omega_B\,a^+\]
involving the creation $a^+$ and annihilation $ a$ operators  with the energy parameter $\omega_B=v\sqrt{2|e|\hbar B/c}=36.2\sqrt{B \text{ [ Tesla]} }$ meV\,. 

 \emph{ For graphene,} one seeks the eigenfunction of Hamiltonian  (\ref{ham0}) in the form 
\begin{equation}
\psi_{sn}^{\alpha}(x)=
\left\{\begin{array}{c}
 C^{1}_{sn}\varphi_{n-1}(x)\\
 C^{2}_{sn}\varphi_{n}(x)\,
\end{array}\right.\,,\label{funcg}
\end{equation}
where   $\varphi_{n}(x)$ are 
orthonormal Hermitian  polynomials with the  numbers $n\ge0$, including  the exponential factor  with one of the space coordinates in the Landau gauge. The number $s=1,2$ numerates the solutions at given $n$.
We  write only one of two $x,y$ space coordinates. The  degeneracy proportional to the magnetic field will be included  in the final results. Every row of the matrix (\ref{ham0}) turns out proportional to the definite  Hermitian polynomial which can be canceled from the eigenvalue equations. We obtain a system of  linear equations 
\begin{equation}
\left(
\begin{array}{cc}
-\varepsilon     & \omega_B\sqrt{n}\\
\omega_B\sqrt{n} & -\varepsilon     
\end{array}%
\right) \times\left\{\begin{array}{c}
C^{1}_{sn}\\
C^{2}_{sn}
\end{array} \right.=0\,  \label{ham1}
\end{equation}
for the eigenvector ${\bf C}_{sn}$ with two eigenvalues, $s=1,2$, 
 \begin{equation}\varepsilon_{sn}=\mp \omega_B\sqrt{ n}\,
 \label{g}\end{equation}
 at given  $ n=1, 2...$
 
 The wave function columns write
\begin{equation}
\begin{array}{c}
C^{1}_{sn}\\
C^{2}_{sn}
\end{array} = \frac{1}{\sqrt{2}}\left\{\begin{array}{c}\, 
1 \\
-1  \end{array}\quad \text{and}\quad \begin{array}{c} 1\\1 \end{array}\,\right.\label{hamu2}
\end{equation}
for  $s=1,2$, correspondingly.
  If $n=0$, there is only one level $\varepsilon_{10}=0$ with $C^1_{10}=0, C^2_{10}=1 $ as follows from Eqs. (\ref{funcg}), (\ref{ham1}).

\begin{figure}[]
\resizebox{.6\textwidth}{!}{\includegraphics{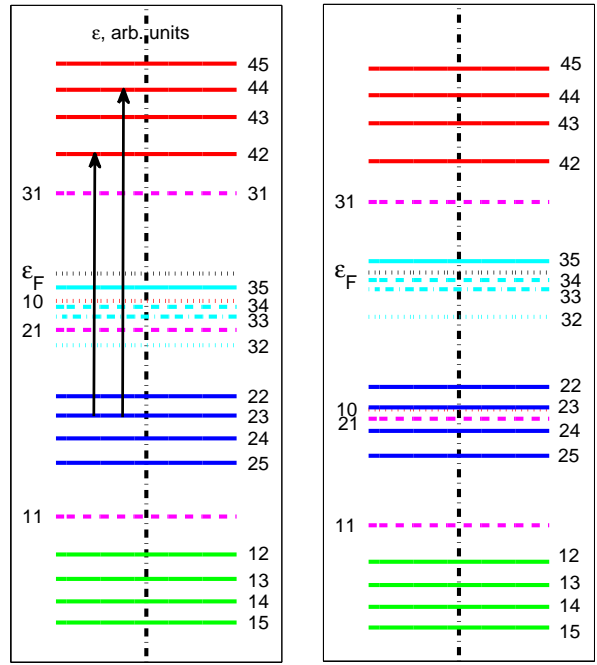}}
\caption{Landau levels in doped bilayer graphene with the Fermi energy $\varepsilon_F=120$ meV in the magnetic field $B=10$ T  and the gate voltage $U=\pm100$ meV at the $K, K'$ points of the Brillouin points  (right and left panels, respectively); the arrows show the electron transitions from the $|23\rangle$ level, where $n=3$ is the Landau number and $s=$2 is the level number  at given $n$ from the bottom.} \label{bild}
\end{figure} 

 \emph{For bilayer graphene,}  we seek the eigenfunction of Hamiltonian  (\ref{hamg}) as a column 
\begin{equation}
\psi_{sn}^{\alpha}(x)=
\left\{\begin{array}{c}
 C^{1}_{sn}\varphi_{n-1}(x)\\
 C^{2}_{sn}\varphi_{n}(x)\\
 C^{3}_{sn}\varphi_{n-1}(x)\\
 C^{4}_{sn}\varphi_{n-2}(x)\,
\end{array}\right.\,,\label{func}
\end{equation}
similar to Eq. (\ref{funcg}).
 The every row
in  Hamiltonian (\ref{hamg}) becomes again proportional to the definite Hermitian function, if the terms with the trigonal warping $\gamma_3$ are omitted. These terms can be considered within the perturbation theory or the semiclassical approximation. 

Canceling  the Hermitian functions from the equations,  we obtain 
a system of the linear equations for the eigenvector~${\bf C}_{sn}$
\begin{equation}
\left(
\begin{array}{cccc}
U-\varepsilon     & \omega_B\sqrt{n} & \gamma_1  & \omega_4\sqrt{n-1}\\
\omega_B\sqrt{n} & U-\varepsilon      & \omega_4\sqrt{n}& 0\\
\gamma_1      &\omega_4\sqrt{n}& -U-\varepsilon  &\omega_B\sqrt{n-1}\\
0\omega_4\sqrt{n-1}& 0 &\omega_B\sqrt{n-1} &-U-\varepsilon
\end{array}%
\right)\left\{\begin{array}{c}
C^{1}_{sn}\\
C^{2}_{sn}\\
C^{3}_{sn}\\
C^{4}_{sn}
\end{array} \right.=0\,  \label{ham2}
\end{equation}
where the band number $s=1,2,3,4$ numerates the solutions at given $n$  from the bottom,  $\omega_B=v\sqrt{2|e|\hbar B/c}\,$ and $\omega_4=\gamma_4\omega_B/\gamma_0$.

The eigenvalues of the matrix in Eq. (\ref{ham2}), see Fig. \ref{bild}, are easily found using the personal computer. Without $\gamma_4$, we have for $\varepsilon_{sn}$ the simplified equation
\[[(U-\varepsilon_{sn})^2- \omega_B^2n][(U+\varepsilon_{sn})^2- \omega_B^2(n-1)]+\gamma_1^2(U^2-\varepsilon_{sn}^2)=0\,.\]
 
 For each Landau number  $n\ge 2$, there are four eigenvalues
 $\varepsilon_{sn}$  and four corresponding eigenvectors, Eq.  (\ref{func}), marked by  the band subscript  $s$;    we will  also use the notation $|sn\rangle$ for levels.
  In addition, there are four  levels. One of them exists for $n=0$ with the eigenvector ${\bf C}_0=(0,1,0,0)$ as is evident from Eqs. (\ref{func}), (\ref{ham2}). Using the perturbation theory, we take the trigonal warping into account \cite{Fa} and  obtain \begin{equation}\varepsilon_1(n=0)=U+\left(\frac{\omega_B\gamma_3}{\gamma_0}\right)^2\sum\limits_{s'}\frac{|C^4_{s'3}|^2}
{U-\varepsilon_{s'}(3)}\,.\end{equation}
 
   Other  three levels  indicated with $n=1$  and $s=1,2,3$ are determined by the first three equations of the system (\ref{ham2}) with $C^4_{s1}=0$.
 The  $|21\rangle$ level is very close to the $|10\rangle$ level. 
 
 \begin{figure}[]
\resizebox{.6\textwidth}{!}{\includegraphics{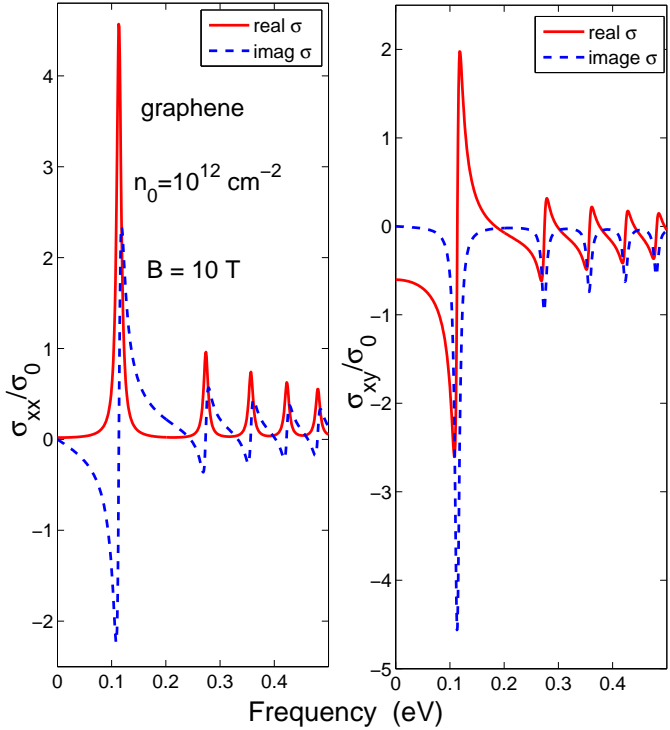}}
\caption{Longitudinal ($\sigma_{xx}$, the left panel)  and Hall ($\sigma_{xy}$, the right panel) dynamical conductivities in the real and imaginary parts for doped graphene with the Fermi energy $\varepsilon_F=120$ meV  in the magnetic field $B=10$ T.}\label{graph}
\end{figure}
 Alternatively,  the semiclassical quantization can be applied for relatively weak magnetic fields when the cyclotron frequency $\frac{d\varepsilon_{sn}}{dn}$ is small compared to the Fermi energy.
 Then, we get the Bohr--Zommerfeld condition in the form
\begin{equation}
\frac{c}{e\hbar B}S(\varepsilon)= 2\pi\left[n+\frac{\mathcal T}{4}+\delta(\varepsilon)\right]\,.
\label{on}\end{equation}
Here $S(\varepsilon)$ is the cross-section area of the classical electron orbit in the $p_x, p_y$ space for the energy $\varepsilon$; $n$ is an integer supposed to be large. The integer $\mathcal T$ is the number of the smooth turning points on the electron orbit. For instance, there are two smooth turning  points for the Landau quadratic model and only one for the skipping electrons reflected by the hard edge.

 If   spin is neglected,  $\delta=0$ and $\mathcal T=2$ for the Landau  quadratic model, and $\delta=1/2$ and $\mathcal T=2$ for
monolayer graphene. In these two cases, the semiclassical  result
coincides  with the rigorous quantization.
 
 Notice, that the  $\delta(\varepsilon)$-phase depends on the energy and it is closely connected
with the topological Berry phase \cite{Be}. The $\delta$-phase was evaluated for bismuth in  Ref. \cite{Falko},  and it was considered again for bismuth in Ref. \cite{MS}. For graphite,   semiclassical quantization was applied in Ref. \cite{Dr}. However, in the general case, the evaluation of the $\delta-$phase  still attracts a widespread interest \cite{TA,CU,KEM,PM,PS,LBM}.

In the simplest case of bilayer graphene without trigonal  warping, we find  the Berry phase \cite{OF}
\begin{equation}\nonumber
\delta(\varepsilon)=\frac{-\varepsilon U}{q^2-\varepsilon^2-U^2}=
\frac{-\varepsilon U}{\sqrt{4U^2\varepsilon^2+(\varepsilon^2-U^2)\gamma_1^2}}.
\label{del1}\end{equation}

For the ungaped bilayer, $U=0$, the Berry phase $\delta(\varepsilon)=0$. The Berry phase
depends on the energy and $\delta=\mp 1/2$ at $\varepsilon =\pm U$, correspondingly. At the larger energy, $\varepsilon\gg U$, the Berry phase $\delta\rightarrow \mp U/\gamma_1$. The effect of the trigonal warping is considered in Ref. \cite{OF}.  

\section{Magneto-optics  in graphene layers}  

 An important peculiarity of conductivity in   magnetic fields is an appearance  of the Hall component $\sigma_{xy}(\omega)$.
 The Hall conductivity violates the rotation symmetry  of graphene around the major $z-$axis. This implies  the rotation of the linear polarized electromagnetic wave, i. e.,  the Faraday and Kerr effects for transmitted and reflected waves, correspondingly.  
 
  First of all, the electron transitions are possible between the levels with the neighboring Landau numbers $n$ and various bands $s$, and therefore the resonance denominators $\Delta_{ss'n}=\varepsilon_{sn}-\varepsilon_{s', n+1}$ arise in the conductivity tensor.

Calculations  \cite{Fa} give the conductivities for layer graphene in the collisionless limit, when the electron collision frequency $\Gamma$ is much less than the level splitting, 
\begin{equation}
\begin{array}{c}
\left.\begin{array}{c} \sigma_{xx}(\omega)\\ i\sigma_{xy}(\omega)
\end{array}\right\}=i{\displaystyle\sigma_0
\frac{4\omega_B^2}{\pi^2}}
{\displaystyle\sum_{n,s,s'}\frac{\Delta f_{ss'n}}{\Delta{ss'n}}|d_{ss'n}|^2}\\ 
\times
\left[(\omega+i\Gamma
+\Delta_{ss'n})^{-1}\pm
(\omega+i\Gamma-\Delta_{ss'n})^{-1} \right]
\,.
\end{array}
\label{dc1}\end{equation}
Here $\Delta f_{ss'n}=f(\varepsilon_{s'n+1})-f(\varepsilon_{sn})$ is the difference of the Fermi functions $f(\varepsilon_{sn})=
[\exp(\frac{\varepsilon_{sn}-\mu}{T})+1]^{-1}$ and  
 \begin{equation}\begin{array}{c}
d_{ss'n}=C^2_{sn}C^{1}_{s'n+1}+C^{3}_{sn}C^{4}_{s'n+1}\nonumber\\
+(C^1_{sn}C^{4}_{s'n+1}+C^{2}_{sn}C^{3}_{s'n+1})\gamma_4/\gamma_0 \end{array}\label{dip}\end{equation} 
is the dipole matrix element expressed in terms of wave functions ${\bf C}_{sn}$, Eqs. (\ref{hamu2}), (\ref{func}),
 and   
$\sigma_0=e^2/4\hbar $
 is the graphene universal conductivity. 

 The electron transitions  obey the
the selection rule
$$\Delta n=1\,.$$

\begin{figure}[]
\resizebox{.6\textwidth}{!}{\includegraphics{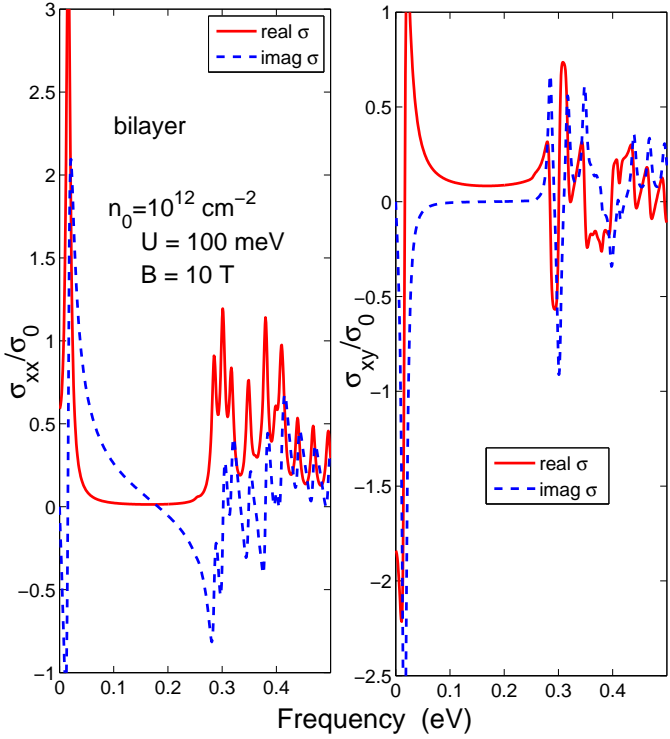}}
\caption{Contribution of  $K$ point in dynamical conductivities for doped bilayer graphene with the Fermi energy $\varepsilon_F=120$ meV  in the magnetic field $B=10$ T;  the interlayer hoping integral $\gamma_1=$360 meV.}\label{bilconK}
\end{figure}
\begin{figure}[]
\resizebox{.6\textwidth}{!}{\includegraphics{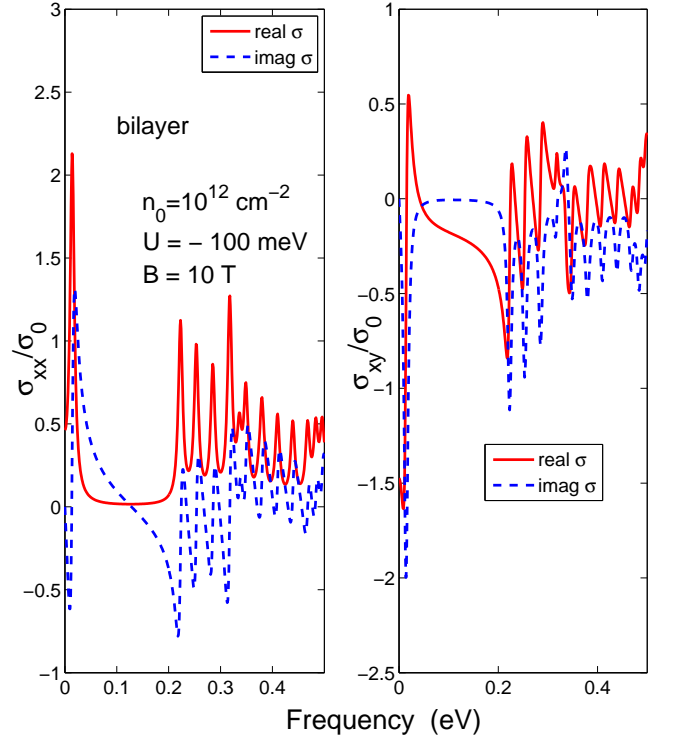}}
\caption{Contribution of  $K'$ point in dynamical conductivities for doped bilayer graphene; the  parameters are the same as in Fig. \ref{bilconK}.}\label{bilconK'}
\end{figure}
Besides,  the renormalization of the dipole moments due to trigonal warping results in weak lines with the selection rule
 $$\Delta n=2.$$ Then, we get an additional term in
Eq. (\ref{dc1}) by substituting 
$$d_{ss'n}= (\tilde{\gamma}_3/\gamma_0)C^2_{sn}C^{4}_{s'n+2}$$
as the matrix element and replacing the subscript
$n+1\rightarrow n+2$.
We have to notice, that the $\gamma_4$ corrections give the linear (in the small parameter $\gamma_4/\gamma_0$) contribution  to the conductivities at the main electron transitions with $\Delta n =1$. The $\gamma_3$ corrections are quadratic, however, they result in an appearance of  new weak resonant transitions with  $\Delta n =2$.

\emph {For graphene,} the calculated conductivities  are shown in Fig. \ref{graph} in the frequency range  0$\div$0.5 eV at the magnetic field $B=$10 T. The electron parameters are taken as follows: the Fermi energy $\varepsilon_F =$120 meV and the electron scattering rate $\Gamma=$5 meV.   The peaks in absorption, the left panel in Fig. \ref{graph}, correspond to the electron transitions  between the levels. The strongest line at 113 meV is exited by the $|10\rangle\rightarrow|21\rangle$ transitions. Other  lines are doublet exited by transitions of the type $|1n\rangle\rightarrow|2,n+1\rangle$ and $|1,n+1\rangle\rightarrow|2,n\rangle$ for $n$ from 1 to 4. All lines obey the selection rule $\Delta n=1$. 






\emph { For bilayer graphene,} the system  of lines is more complicated. At $K$ point (the gate voltage $U$=100 meV), it is shown in Fig. \ref{bild} schematically  and in Table for the Fermi energy $\varepsilon_F = 120$ meV. The contributions of  $K$ and $K'$ points into conductivities are compared in Figs. \ref{bilconK} and \ref{bilconK'}.

\begin{table}[]
\caption{\label{tb1} Landau levels $|sn\rangle$  for $2\le n\le $12 in bilayer; other levels are
$|10\rangle$= 100, $|11\rangle$= -384, $|21\rangle$= 82, $|31\rangle$= 401 in meV.  }
                \begin{tabular}{|c|c|c|c|c|c|c|c|c|c|c|c|}\hline
 \backslashbox{$s$}{$n$}  &2     & 3    & 4   & 5    & 6    & 7    &8     & 9& 10&11&12\\
 \hline
 1 & -417 & -447 & -472& -496 & -517 & -537 & -556 &-574&-592&-608&-624\\
   2& -94& -99&-110&-124&-140&-155&-172&-186&-202&-216&-231\\ 
3& 79& 86&98&113&130&146&162&178 &193&208&223\\ 
4& 432& 460&484&507&527&547&565&583 &600&616&631\\
\hline
\end{tabular}
\end{table}





The absorption resonances in Fig. \ref{bilconK}, the left panel, are composed by the following electron transitions:

16 meV: $|35\rangle\rightarrow |36\rangle$, 

254 meV (very weak):$|25\rangle\rightarrow |36\rangle$, 

285 meV: $|26\rangle\rightarrow |37\rangle, |27\rangle\rightarrow |36\rangle$, 

301 meV: $ |10\rangle\rightarrow |31\rangle$,

317 meV: $|27\rangle\rightarrow |38\rangle, |28>\rightarrow |37\rangle$, 

348 meV: $|28\rangle\rightarrow |39\rangle, |29\rangle\rightarrow |38\rangle, |33\rangle\rightarrow|42\rangle,$ $|21\rangle\rightarrow |42\rangle$,

380 meV: $|210\rangle\rightarrow |39\rangle, |29\rangle\rightarrow |310\rangle, |32\rangle\rightarrow |43\rangle$,

398  meV (weak): $|33\rangle\rightarrow |44\rangle$,

410  meV: $|211\rangle\rightarrow |310\rangle, |210\rangle\rightarrow |311\rangle, |35\rangle\rightarrow |46\rangle, |34\rangle\rightarrow |45\rangle, |35>\rightarrow |46\rangle$,

439 meV: $|212\rangle\rightarrow |311\rangle, |211\rangle\rightarrow |312\rangle$,

468  meV: $|213\rangle\rightarrow |312\rangle, |212\rangle\rightarrow |313\rangle$,

496 meV: $|214\rangle\rightarrow |313\rangle, |213\rangle\rightarrow |314\rangle, |22\rangle\rightarrow |31\rangle$.
\begin{figure}[]
\resizebox{.6\textwidth}{!}{\includegraphics{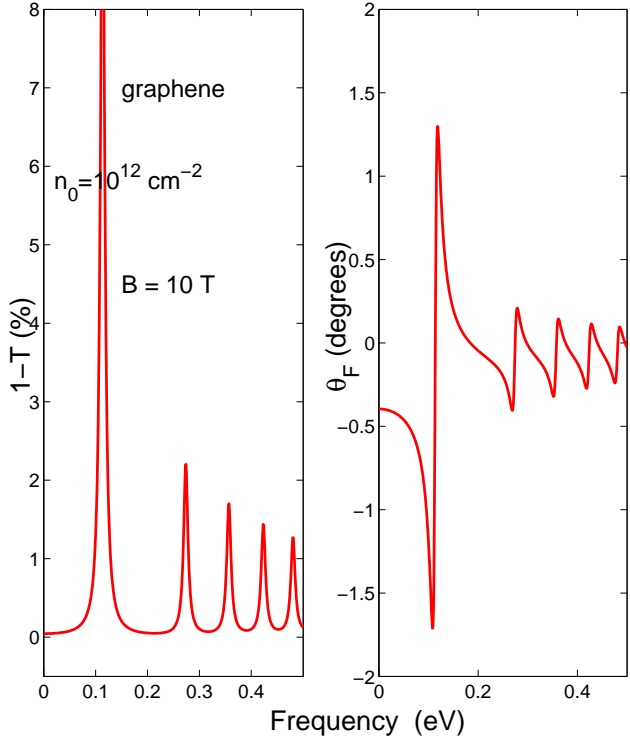}}
\caption{Transmittance  ($T$, the left panel)  and Faraday rotation angle ($\Theta_F$, the right panel) for doped graphene with the Fermi energy $\varepsilon_F=120$ meV  in the magnetic field $B=10$ T.}\label{trfargr}
\end{figure}

Graphene and bilayer graphene affect \emph {the transmission and the Faraday rotation} to the linear order in the fine structure constant  $\alpha$ as well as the reflected light intensity  quadratic in $\alpha$. The conductivities 
$\sigma_{xx}(\omega)$ and $\sigma_{xy}(\omega)$ allow  calculating the Faraday rotation and the transmittance  as functions of the frequency. Because the conductivity of the layers are small, we can use   the linear approximation in $\alpha$. The transmission coefficient $T$ and the Faraday angle $\Theta_F$ for the free standing layers  write  as  
\begin{equation}
T = 1-\frac{4\pi}{c}\text{Re}\,\sigma_{xx}(\omega)\,,
\Theta_F= \frac{2\pi}{c}\text{Re}\,\sigma_{xy}(\omega)\,.
\label{trbi}\end{equation}

\begin{figure}[]
\resizebox{.6\textwidth}{!}{\includegraphics{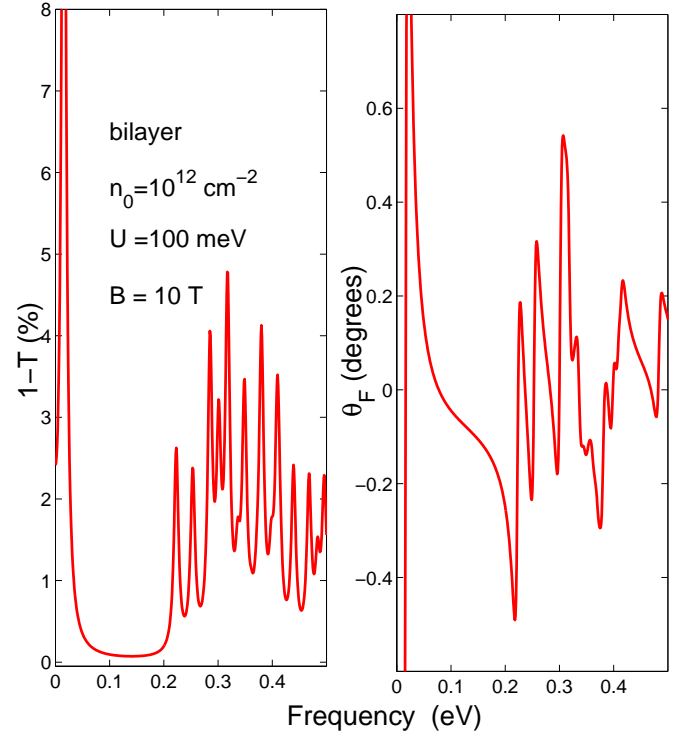}}
\caption{Transmittance  ($T$, the left panel)  and Faraday rotation angle ($\Theta_F$, the right panel) for doped bilayer with the Fermi energy $\varepsilon_F=120$ meV  in the magnetic field $B=10$ T.}\label{trfarbi}
\end{figure}
 
 Results of  calculations are shown in Fig. \ref{trfargr} for graphene and in Fig. \ref{trfarbi} for bilayer. We take into account that the points
 $K$ and $K'$ have the different electron levels in bilayer  at $U\neq 0$, and both these points contribute independently in observable quantities.

It is evident that the interpretation of the Faraday rotation governed by
the conductivity $\sigma_{xy}(\omega)$  is much more complicated
  in comparison with the transmittance controlled by the longitudinal conductivity.  
 
 The positions of the lines  for  fields in the range of 10 -- 30~ T agree   with observations of Refs. \cite{OFS,CBN}.
 
 \section{Effect of temperatures and magnetic fields on the chemical potential}   

In the previous figures, we assume that the temperature and  level width $\Gamma$ are much less than the level splitting $|\varepsilon_{s, n+1}-\varepsilon_{s, n}|$. As is known, the chemical potential $\mu$ in semiconductors
changes while the temperature increases. We know also that  de Haas--van Alphen oscillations can be observed at low temperatures. We can find the amplitude of these effects
assuming that the carrier concentration $n_0$ should be constant when the temperature or the magnetic field  rises. 

For simplicity reasons, let us consider now graphene, where  $\varepsilon_{sn}=\mp \omega_B\sqrt{n}$,  $ n=0,1,2..,$ $s=1,2$, $g=4$, $dn/d\varepsilon_s=2\varepsilon_s/\omega_B^2$, $\omega_B=v\sqrt{2e\hbar B/c}$.  Electrons in the upper band and holes in the lower band influence each other, especially at high temperatures. In order to incorporate electrons ($\mu>0$) and holes ($\mu<0$) in a single scheme, let us introduce the variable $\varepsilon_n=\pm \varepsilon_{sn}$ for $ s=2,1$ correspondingly. Thus 
the carrier concentrations for 2d systems write
\begin{equation}\nonumber
n_0=\text{sign}(\mu)\frac{geB}{2\pi\hbar c} \sum_{n=0}^{\infty}[f(\varepsilon_{n}-\mu)-f(\varepsilon_{n}+\mu)]
\label{conc}\end{equation}
where  $f(\varepsilon_{n}-\mu) $ is the Fermi-Dirac function,
 $g$ is  the spin and valley factor, and $n_0$ is considered as positive whereas $\mu$ can be of  both signs.

When the level splitting is far less than the chemical potential, we can use  the Poison summation formula
\begin{equation}
n_0=\frac{geB}{2\pi\hbar c}\left[\int_{n=0}^{\infty} dn y(\varepsilon_{n})
+\sum_{k\neq0}\int_{n=0}^{\infty} dn e^{2\pi ikn}y(\varepsilon_{n})\right]\,,\label{conc1}\end{equation}
with $y(\varepsilon_{n})=|f(\varepsilon_{n}-\mu)-f(\varepsilon_{n}+\mu)|$.
The second term in brackets goes to zero at $B\rightarrow 0$ and then the first term gives the carrier concentration without the magnetic field
$$n_0(B\rightarrow 0)=\frac{geB}{2\pi\hbar c}\int_0^{\infty} d\varepsilon dn/d\varepsilon y(\varepsilon).$$
 This term depends on the magnetic field only through the chemical potential:
\begin{equation}
n_0(B\rightarrow 0)=\frac{g}{2\pi v^2\hbar^2 }\int_0^{\infty} \varepsilon d\varepsilon y(\varepsilon)\,.
\label{conc3}\end{equation}

\begin{figure}[]
\resizebox{.4\textwidth}{!}{\includegraphics{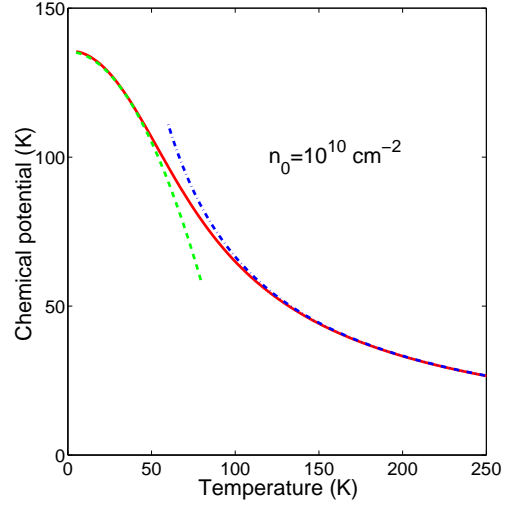}}
\caption{Chemical potential versus temperature for the carrier concentration $10^{10}$ cm$^{-2}$; the exact solution to Eq. (\ref{conc3}) in solid line, asymptotes for low, Eq. (\ref{conc44}), and high, Eq. (\ref{high}),  temperatures (dashed and dashed-dotted lines, correspondingly). }\label{chem}
\end{figure}

At given  carrier concentration $n_0$ determined  by doping or a  bias voltage, the equation (\ref{conc3}) gives the temperature dependence   of the chemical potential (see Fig. \ref{chem}) and, particularly, the Fermi energy at zero temperature, $\mu(T=0)\equiv\varepsilon_F$:
$$\varepsilon_F=\mp \hbar v \sqrt{4\pi n_0/g},$$
where $\mp$ stand for holes and electrons correspondingly.

We  find the temperature dependence of the chemical potential  differentiating the equation (\ref{conc3}) with respect to the temperature:
\begin{equation}
0=\int_0^{\infty} \varepsilon d\varepsilon \left[\frac {\partial y(\varepsilon)}{\partial \mu}\frac{d\mu}{dT}+\frac{\partial y(\varepsilon)}{\partial T}\right].
\label{conc4}\end{equation}

At low temperatures, $T\ll|\mu|$, we can evaluate these integrals. Because the integrand in the first integral is proportional to the Dirac $\delta-$function,
it gives $\mu \frac{d\mu}{dT}$. The second integral  equals $\pi^2 T/3$.
Then, we get the equation
$$\mu\frac{d\mu}{dT}+\frac{\pi^2}{3}T=0$$
with a solution
$$\mu^2=\varepsilon_F^2-\frac{\pi^2}{3}T^2$$
or at low temperatures \begin{equation}\mu=\varepsilon_F-\frac{\pi^2}{6}T^2/\varepsilon_F\,.
\label{conc44}\end{equation}

At high temperatures, $T\gg |\mu|$, the  integral, Eq. (\ref{conc3}), gives the approximate dependence
\begin{equation}|\mu|=\frac{\pi n_0 (\hbar v)^2}{4\ln{2}\, T}\label{high}\end{equation}
shown in Fig. \ref{chem}.
While the temperature grows, the chemical potential tends to its value $\mu=0$ in  undoped graphene.

The second term in brackets (\ref{conc1}) is easily calculated at low temperatures:
\begin{equation}\begin{array}{c}
I(B)=
{\displaystyle \sum_{k\neq0}\int_0^{\infty} d\varepsilon \varepsilon e^{2\pi ikn}y(\varepsilon)}\\=
{\displaystyle 2\pi |\mu| T\sum_{k>0}\frac{\sin(2\pi k n_{\mu})}{\sinh(2\pi^2kn'_{\mu}T)}},
\end{array}\label{conc5}\end{equation}
where $n_{\mu}=\mu^2/\omega_B^2$ is the Landau number, corresponding to the Fermi energy and $n'_{\mu}=2\mu/\omega_B^2$ is its derivative.
Owing to the denominator, this term is relatively small at weak magnetic fields, $4\pi^2|\mu| T \gg\omega^2_B$, and we accept this condition. 

Then, we consider the effect of the magnetic field on the chemical potential in the way similar to Eq. (\ref{conc4}):
\begin{equation}
0=\frac{\partial \mu}{\partial B}\int_0^{\infty} \varepsilon d\varepsilon \frac {\partial y(\varepsilon)}{\partial \mu}+\frac{\partial I(B)}{\partial B}.
\label{conc6}\end{equation}
The integral here equals  $|\mu|$. Integrating the obtained differential equation  with  the initial  condition $\mu=\varepsilon_F$ at $B=T=0$ and taking Eq. (\ref{conc44}) into account, we get the  dependence of the chemical potential on the magnetic field and the temperature
\begin{equation}\mu=\varepsilon_F-I(B)/\varepsilon_F-\frac{\pi^2}{6}T^2/\varepsilon_F,
\label{conc7}\end{equation}
where we have to take $\varepsilon_F$ instead of $\mu$ in $I(B)$ (\ref{conc5}) for the case of the relatively large $|\varepsilon_F|\gg\omega_B$.

The equation (\ref{conc7}) represents   de Haas--van Alphen oscillations and the  chemical potential  shift in 2d systems. We emphasize that the considered effects have no relations to the electron-electron  or electron-phonon interactions but come out purely from the broadening of the distribution function. The corresponding  impact, proportional to the temperature, on the width $\Gamma$ is considered by Ozerin in Ref. \cite{Oz}.
 
 \section{Summary and conclusions}

In conclusions, we discuss the dynamic conductivity of monolayer and bilayer graphene in the optical range, where the ac frequency is much larger than the electron scattering rate.   The trigonal warping in bilayer can be considered within the perturbation theory at  strong magnetic fields larger than 1 T approximately. For weak magnetic fields, when the Fermi energy  much larger than the cyclotron frequency, the semiclassical quantization with the Berry phase included can be applied. The main electron transitions obey the selection rule $\Delta n=1$ for the Landau number $n$, however  $\Delta n=2$ transitions due to the trigonal warping with the small probability are also possible. The SWMC  parameters are used in the fit  taking  their values from the previous dHvA measurements.   The calculated conductivities, longitudinal and Hall's,
permit to evaluate transmittance and Faraday's rotation of light in the graphene layers. The agreement between the calculations and the measured  Faraday rotation and transmissions in graphene in the quantizing magnetic fields is achieved. Assuming that the carrier concentrations is fixed by gate voltage or  dopant, we find the effect of temperatures and magnetic fields on the chemical potential.  

\acknowledgments
This work was supported by the Russian Foundation for Basic
Research and the SIMTECH Program, New Centure of Superconductivity: Ideas, Materials and Technologies (grant no. 246937).

\end{document}